\documentclass[12pt]{article}
\begin{document}
\newfont{\elevenmib}{cmmib10 scaled\magstep1}%
\renewcommand{\theequation}{\arabic{section}.\arabic{equation}}
\newcommand{\tabtopsp}[1]{\vbox{\vbox to#1{}\vbox to12pt{}}}
\font\larl=cmr10 at 24pt
\newcommand{\es}{\got s}

\newcommand{\preprint}{
            \begin{flushleft}
   \elevenmib Yukawa\, Institute\, Kyoto\\
            \end{flushleft}\vspace{-1.3cm}
            \begin{flushright}\normalsize  \sf
            YITP-01-34\\
           {\tt hep-th/0105164} \\ May 2001
            \end{flushright}}
\newcommand{\Title}[1]{{\baselineskip=26pt \begin{center}
            \Large   \bf #1 \\ \ \\ \end{center}}}
\hspace*{2.13cm}%
\hspace*{0.7cm}%
\newcommand{\Author}{\begin{center}\large \bf
           V.\,I.\, Inozemtsev\footnote{
permanent address: BLTP JINR, 141980 Dubna, Moscow Region, Russia}
 and R.\, Sasaki \end{center}}
\newcommand{\Address}{\begin{center}
            Yukawa Institute for Theoretical Physics\\
     Kyoto University, Kyoto 606-8502, Japan
      \end{center}}
\newcommand{\Accepted}[1]{\begin{center}{\large \sf #1}\\
            \vspace{1mm}{\small \sf Accepted for Publication}
            \end{center}}
\baselineskip=20pt

\preprint
\thispagestyle{empty}
\bigskip
\bigskip
\bigskip

\Title{Universal Lax pairs for Spin Calogero-Moser Models and Spin
Exchange Models}
\Author

\Address
\vspace{1cm}

\begin{abstract}
For any root system $\Delta$ and an  irreducible representation
${\cal R}$ of the reflection (Weyl) group $G_\Delta$ generated by
$\Delta$, a {\em spin Calogero-Moser model\/} can be defined for each of the
potentials: rational, hyperbolic, trigonometric and elliptic.
For each member $\mu$ of ${\cal R}$, to be called a ``site", we associate
a vector space ${\bf V}_{\!\mu}$ whose element is called a ``spin".
Its dynamical variables are the canonical coordinates $\{q_j,p_j\}$
of a particle
in ${\bf R}^r$,  ($r=$ rank of
$\Delta$), and spin exchange operators $\{\hat{\cal P}_\rho\}$
($\rho\in\Delta$) which
exchange the spins at the sites $\mu$ and $s_{\rho}(\mu)$.
Here $s_\rho$ is the reflection generated by $\rho$.
For each $\Delta$ and ${\cal R}$ a {\em spin exchange model\/} can be
defined.
The Hamiltonian  of a spin exchange model is a linear combination of the
spin exchange operators only. It is obtained by ``freezing" the canonical
variables at the  equilibrium point of the corresponding classical
Calogero-Moser model. For $\Delta=A_r$ and ${\cal R}=$ vector representation
it reduces to the well-known Haldane-Shastry model.
Universal Lax pair operators for both spin Calogero-Moser models and spin
exchange models are presented which enable us to construct
as many conserved quantities as the number of sites for {\em degenerate\/}
potentials.
\end{abstract}
\bigskip
\bigskip
\bigskip

\section{Introduction}
\label{intro}
\setcounter{equation}{0}

The essential part of our knowledge of quantum many-body systems is
concerned
with integrable models in one dimension. Among them, the
Calogero-Moser models
\cite{Cal,Sut,CalMo,Ino0} with long-range interactions are
most popular during last
decade. Their links to the models of solid-state physics
\cite{halsha,ino1, suthsha2,fmp,HikWa,simal,ber,ino2} have been found,
and they are based on the possibility to introduce also the spin exchange
interaction in a translation-invariant form. However, the CM models can be
formulated  n classical
and quantum mechanics for any root system
\cite{OP1,DHoker_Phong,bcs1,bcs2,bms,kps}, and one can guess that
introduction of spin exchange can be done at least for some root systems
too.
There
were several attempts \cite{fmp,HikWa,simal,yam} in this direction,
but they were far
from being universal in a way for introducing spin into the CM models.

In this paper, we consider the possibility of unifying all the previous
approaches to spin Calogero-Moser models and related models of spin exchange
interactions obtained by ``freezing" the canonical variables at the
equilibrium points of the corresponding classical CM systems.
This can be done by constructing universal Lax representations for
degenerate
forms of the CM
potentials. There are also some indications that the corresponding models
with most general elliptic potentials are also integrable \cite{ino1,ino2},
but the
construction of Lax pair in this case does not lead directly to
integrability.

The organization of the paper is as follows. In Section 2, the universal Lax
operators for the CM models with degenerate potentials \cite{bcs2,bms}
is briefly
recapitulated. The way of introducing spin exchange in the framework
of the above
formalism is proposed in Section 3 so as to prove the integrability
of the spin CM
models for all root systems.
The existence of conserved quantities is guaranteed by
the  ``sum to zero" condition for the second Lax operator. Section 4 is
devoted to the models with spin exchange operators only. The
corresponding Lax operators lead in the trigonometric case and $A_{r}$
root system to  Haldane-Shastry model \cite{halsha}.
The Polychronakos model \cite{fmp,yam} corresponds in this
approach to the rational case with a confining $q^2$ potential.
The final section is devoted to summary and comments.
\section{Universal Lax Operator for Calogero-Moser Model with Degenerate
Potential}
\label{Lax}
\setcounter{equation}{0}

In this section we briefly recapitulate the essence of Calogero-Moser models
based on any root system $\Delta$ (applicable to the exceptional and
non-crystallographic root system) and the associated universal
Lax pair formalism
along with appropriate notation \cite{bcs1,bcs2,bms,kps} and background
\cite{OP1,DHoker_Phong} for the main body of this paper. Those who are
familiar
with the universal Lax pair formulation may skip this section and return
when
necessity arises. A   Calogero-Moser model is a
Hamiltonian system associated with a root system $\Delta$
of rank \(r\), which is a set of
vectors in $\mathbf{R}^{r}$ with its standard inner product,
invariant under reflections
in the hyperplane perpendicular to each
vector in $\Delta$.  In other words,
\begin{equation}
   s_{\alpha}(\beta)\in\Delta,\quad\forall \alpha,\beta\in\Delta,
\quad s_{\alpha}(\beta)=\beta-(\alpha^{\vee}\!\!\cdot\beta)\alpha,
\quad \alpha^{\vee}\equiv 2\alpha/|\alpha|^{2}.
\end{equation}
The set of reflections $\{s_{\alpha},\,\alpha\in\Delta\}$ generates a
group $G_{\Delta}$, a finite reflection group, known as the Coxeter
(Weyl) group.
The set of roots $\Delta$ is decomposed into a disjoint sum
of the positive roots $\Delta_+$ and negative roots $\Delta_-$.
The dynamical variables of the Calogero-Moser model are the coordinates
$\{q_{j}\}$ and their canonically conjugate momenta $\{p_{j}\}$,which will
be
denoted by vectors in $\mathbf{R}^{r}$ with the standard inner product:
\begin{equation}
   q=(q_{1},\ldots,q_{r}),\qquad p=(p_{1},\ldots,p_{r}),\qquad
p^2=p\cdot p=\sum_{j=1}^rp_j^2.
\end{equation}
The Hamiltonian for  classical
Calogero-Moser model with a degenerate potential reads:
\begin{equation}
   \label{cCMHamiltonian}
   \mathcal{H}_C = {1\over 2} p^{2} + {1\over2}\sum_{\rho\in\Delta_+}
   {g_{|\rho|}^{2} |\rho|^{2}}
   \,V(\rho\cdot q),
\end{equation}
in which the potential function $V$ is listed in the following Table 1:
\begin{center}
 \begin{tabular}{|l|c|c|c|}
 \hline
  & $V(u)$ & $x(u)$ & $y(u)$ \\
 \hline
 rational & $1/u^2$ & $1/u$ & -$1/u^2$ \\
 \hline
 trigonometric & $1/\sin^2u$ & $\cot u$ & -$1/\sin^2 u$ \\
 \hline
 hyperbolic & $1/\sinh^2u$ & $\coth u$ & -$1/\sinh^2 u$ \\
 \hline
 \end{tabular}\\
 \bigskip
 Table 1: Functions appearing in the Hamiltonian and Lax pair.
\end{center}
Here we have omitted the scale factor for the trigonometric (hyperbolic)
potential, for simplicity.
The associated universal Lax pair operators read
\begin{eqnarray}
   L &=&  p\cdot\hat{H}+X,\qquad
X=i\sum_{\rho\in\Delta_{+}}g_{|\rho|}
   \,\,(\rho\cdot\hat{H})\,x(\rho\cdot q)\,\hat{s}_{\rho},
   \label{LaxOpDef}\\
   \widetilde{M} &=&
   {i\over2}\sum_{\rho\in\Delta_{+}}g_{|\rho|}|\rho|^2\,y
   (\rho\cdot q)\,\hat{s}_{\rho},
   \label{Mtildef}
\end{eqnarray}
in which the functions $x(u)$ and $y(u)$ are listed in the Table 1.
These functions are related by
\begin{equation}
   y(u)\equiv dx(u)/{du},\quad
    V(u)=-y(u)=x^2(u)+\mbox{constant}.
   \label{hdef}
\end{equation}
The real {\em positive\/} coupling constants \(g_{|\rho|}\)
 are defined on orbits of the corresponding
reflection group, {\it i.e.} they are
identical for roots in the same orbit. That is, for the simple Lie
algebra cases one coupling constant
\(g_{|\rho|}=g\) for all roots in simply-laced models
and  two independent coupling constants, \(g_{|\rho|}=g_L\)
for long roots and \(g_{|\rho|}=g_S\) for
short roots in non-simply laced models.
The operators $\hat{H}_j$ and $\hat{s}_{\rho}$ obey the following
commutation relations
\begin{eqnarray}
   \label{OpAlgebra1}
   [\hat{H}_{j},\hat{H}_{k}]=0, \\
   \label{OpAlgebra2}
   [\hat{H}_{j},\hat{s}_{\alpha}] = \alpha_{j}
   (\alpha^{\vee}\!\!\cdot\hat{H})\hat{s}_{\alpha},
   \\
   \label{OpAlgebra3}
   \hat{s}_{\alpha}\hat{s}_{\beta}\hat{s}_{\alpha}
   =\hat{s}_{s_{\alpha}(\beta)},\quad
\hat{s}_{\alpha}^2=1,\quad \hat{s}_{-\alpha}=\hat{s}_{\alpha}.
\end{eqnarray}
In terms of these commutation relations it is easy to show that
the canonical equations of motion can be represented in an operator form:
\begin{equation}
   \label{LaxEquation}
 \dot{q}_{j}=  p_{j},\quad  \dot{p}_{j}= -{\partial{\cal
H}_C\over{\partial q_{j}}}
\Longleftrightarrow
   {d\over dt}{L}=[L,\widetilde{M}].
\end{equation}

Let us choose an irreducible  $D$-dimensional representation of the
reflection (Weyl) group $G_\Delta$ to be denoted by ${\cal R}$.
It is a collection of ${\bf R}^r$  vectors,
to be called a  ``site",
which form a single Weyl orbit:
\begin{equation}
{\cal R}=\{\mu^{(1)},\ldots,\mu^{(D)}|\mu^{(k)}\in {\bf R}^r\}.
\end{equation}
That is any site of ${\cal R}$ can be obtained from any other site by the
action of the reflection (Weyl) group. Thus the (length)${}^2$ of the
vectors
$\mu^{(k)}$ are equal:
\begin{equation}
(\mu^{(j)})^2=(\mu^{(k)})^2,\quad \forall \mu^{(j)}, \mu^{(k)}\in{\cal R}.
\end{equation}
Then $L$ and $\widetilde{M}$ are $D\times D$ matrices whose elements are
given by
\begin{equation}
(\hat{H}_{j})_{\mu\nu}=\mu_j\delta_{\mu\nu},\quad
(\hat{s}_{\rho})_{\mu\nu}=\delta_{\mu,s_\rho(\nu)}=
   \delta_{\nu,s_\rho(\mu)}.
\end{equation}
The essence of the Lax pair is the following set of identities
among the functions $\{x(\rho\cdot q)\}$ and $\{y(\rho\cdot q)\}$
expressed in matrix forms:
\begin{equation}
  \label{firstiden}
 [X, \widetilde{M}]= -\hat{H}\cdot{\partial{\cal V}\over{\partial q}},
\quad {\cal V}={1\over2}\sum_{\rho\in\Delta_+}
   {g_{|\rho|}^{2} |\rho|^{2}}
   \,V(\rho\cdot q),
\end{equation}
\vspace{-12pt}
\begin{equation}
 [\, p\cdot\hat{H}, \widetilde{M} ]=i[-{1\over2}{\partial^2\over{\partial
q^2}}, X],
\label{seciden}
\end{equation}
in which the right hand side of (\ref{firstiden}) is a diagonal matrix.
The matrix $\widetilde{M}$ has a special property (see (2.36)
of \cite{bms}):
\begin{equation}
   \sum_{\mu\in{\cal R}}\widetilde{M}_{\mu\nu}=
   \sum_{\nu\in{\cal R}}\widetilde{M}_{\mu\nu}=
   -i{\cal V}_S,\quad
{\cal V}_S={1\over2}\sum_{\rho\in\Delta_{+}}g_{|\rho|}|\rho|^2\,V
      (\rho\cdot q),
   \label{Mtilsum}
\end{equation}
in which ${\cal V}_S$ is independent of $\mu$ and $\nu$.
Note that ${\cal V}_S$ is different from ${\cal V}$ in (\ref{firstiden}),
which is quadratic in the coupling $g_{|\rho|}$, whereas ${\cal V}_S$ is
linear.
We can define a new matrix $M$,
\begin{equation}
   M=\widetilde{M}+i{\cal V}_S\times I,\quad I:\ \mbox{Identity operator},
   \label{Mdefnew}
\end{equation}
which satisfies {\em sum up to zero} condition
\begin{equation}
   \sum_{\mu\in{\cal R}}M_{\mu\nu}=
   \sum_{\nu\in{\cal R}}M_{\mu\nu}=0.
   \label{sumMzero}
\end{equation}
Since the elements of the matrices $X$ and $M$ are numbers and ${\cal
V}_S\times I$ commutes with $X$ we have from
 (\ref{firstiden})
\begin{equation}
 [X, M]= -\hat{H}\cdot{\partial{\cal V}\over{\partial q}},
\end{equation}
which is the content of the usual Lax pair.

\section{Spin Calogero-Moser Model with Degenerate
Potential}
\label{spincal}
\setcounter{equation}{0}

Now let us define a spin Calogero-Moser model associated with a
root system $\Delta$ and an irreducible representation ${\cal R}$
of the reflection (Weyl) group $G_\Delta$, that is the set of ``sites".
A dynamical state of the model is a wavefunction $\psi(q)$ times
a vector $\psi_S$ which takes value in the $D$ multiple of a vector
space ${\bf V}$;
\begin{equation}
\psi_S\in {\buildrel D\over\otimes}{\bf V}.
\end{equation}
Each $\bf V$ is associated with site $\mu$. In other words
$\psi_S$ can be represented by its component spin $\psi_S^{(\mu)}$
at the site $\mu$, or $\psi_S^{(j)}$ at site $j$ for short:
\begin{equation}
\psi_S=|\psi_S^{(1)},\ldots,\psi_S^{(D)}>.
\end{equation}
Let us introduce a spin exchange operator $\hat{\cal P}_{\rho}$
associated with each root $\rho\in\Delta$:
\begin{equation}
\hat{\cal P}_{\rho}: \psi_S\to\hat{\cal P}_{\rho}\psi_S,\quad
(\hat{\cal P}_{\rho}\psi_S)^{(\mu)}=\psi_S^{(s_{\rho}(\mu))},
\quad \forall\mu\in{\cal R}.
\end{equation}
Obviously $\{\hat{\cal P}_{\rho}\}$ ($\rho\in\Delta$) satisfy the
same commutation relations as $\{\hat{s}_{\rho}\}$:
\begin{equation}
  \hat{\cal P}_{\alpha}\hat{\cal P}_{\beta}\hat{\cal P}_{\alpha}
   =\hat{\cal P}_{s_{\alpha}(\beta)},\quad
\hat{\cal P}_{\alpha}^2=1,\quad \hat{\cal P}_{-\alpha}=\hat{\cal
P}_{\alpha},
\end{equation}
and $\hat{s}_{\alpha}$, $\hat{H}_j$ and $\hat{\cal P}_{\beta}$
commute since
they act on different spaces
\begin{equation}
[\hat{s}_{\alpha},\hat{\cal P}_{\beta}]=0=[\hat{H}_{j},\hat{\cal
P}_{\beta}].
\end{equation}
Likewise the quantum operators
$\{q_j\}$ and $\{p_k\}$ commute with $\hat{\cal P}_{\rho}$:
\begin{equation}
[q_j,\hat{\cal P}_{\rho}]=0=[p_k,\hat{\cal P}_{\rho}],\quad
j,k=1,\ldots, r,\quad \forall \rho\in\Delta.
\end{equation}

By multiplying $\hat{\cal P}_{\rho}$ to the functions $x(\rho\cdot q)$
and $y(\rho\cdot q)$ in $X$ and $\widetilde{M}$, we define new matrices
$X_S$ and $\widetilde{M}_S$:
\begin{eqnarray}
X_S&=&i\sum_{\rho\in\Delta_{+}}g_{|\rho|}
   \,\,(\rho\cdot\hat{H})\,x(\rho\cdot q)\,\hat{\cal
P}_{\rho}\hat{s}_{\rho},
   \label{XsDef}\\
   \widetilde{M}_S &=&
   {i\over2}\sum_{\rho\in\Delta_{+}}g_{|\rho|}|\rho|^2\,y
   (\rho\cdot q)\,\hat{\cal P}_{\rho}\hat{s}_{\rho},
   \label{Mstildef}
\end{eqnarray}
whose elements are no longer numbers but operators now.
As in the previous section we  define a new matrix $M_S$,
\begin{equation}
   M_S=\widetilde{M}_S+i{\cal A}\times I,
   \label{MSdefnew}
\end{equation}
which satisfies {\em sum up to zero} condition, too
\begin{equation}
   \sum_{\mu\in{\cal R}}({M_S})_{\mu\nu}=
   \sum_{\nu\in{\cal R}}({M_S})_{\mu\nu}=0.
   \label{sumMSzero}
\end{equation}
The operator ${\cal A}$ now depends on the spin exchange
operators $\{\hat{\cal P}_{\rho}\}$:
\begin{equation}
{\cal A}={1\over2}\sum_{\rho\in\Delta_{+}}g_{|\rho|}|\rho|^2\,V
      (\rho\cdot q)\hat{\cal
P}_{\rho}.
\end{equation}
 Since the commutation relations of
$\{\hat{H}_j,\hat{s}_\rho\}$ and
$\{\hat{H}_j,\hat{\bf s}_\rho\equiv\hat{\cal P}_\rho\hat{s}_\rho\}$
are identical we have the following main result
\begin{equation}
  \label{mainiden}
 [X_S, \widetilde{M}_S]= -\hat{H}\cdot{\partial{\cal V}\over{\partial q}},
\end{equation}
in which the right hand side does not contain operators $\{\hat{\cal
P}_{\rho}\}$. This is because they cancel out by the relation
$\hat{\cal
P}_{\rho}^2=1$.
The right hand side can be replaced by the obvious identity in quantum
theory
\begin{equation}
 -\hat{H}\cdot{\partial{\cal V}\over{\partial q}}=i[{\cal H}_C,
p\cdot\hat{H}].
\end{equation}
If we rewrite $\widetilde{M}_S$ in terms of $M_S$, we obtain
\begin{equation}
[X_S, M_S-i{\cal A}]=i[{\cal H}_C, \,p\cdot\hat{H}],
\label{mainiden2}
\end{equation}
in which the second commutator in the left hand side no longer vanishes.
By adding (\ref{mainiden2}) to
\begin{equation}
 [p\cdot\hat{H}, M_S-i{\cal A} ]=i[{p^2\over{2}}, X_S],
\label{seciden2}
\end{equation}
we arrive at the desired equation
\begin{eqnarray}
 [p\cdot\hat{H}+X_S, M_S]&=&i[{\cal H}_S,
\, p\cdot\hat{H}+X_S],
\\
{\cal H}_S&\equiv&{\cal H}_C-{\cal A}=
{1\over 2} p^{2} + {1\over2}\sum_{\rho\in\Delta_+}
   {|\rho|^{2}g_{|\rho|}(g_{|\rho|}-\hat{\cal P}_\rho) }
   \,V(\rho\cdot q),
\label{spinHamdef}
\end{eqnarray}
which is a {\em universal} Lax equation for the spin Calogero-Moser model
\begin{equation}
i[{\cal H}_S,L_S]=[L_S,M_S],\quad
L_S=p\cdot\hat{H}+X_S
\label{spinLax}
\end{equation}
defined by the Hamiltonian ${\cal H}_S$ (\ref{spinHamdef}).
That is, this applies to any spin Calogero-Moser models based on
any root system $\Delta$ and any irreducible representation ${\cal R}$
of the reflection (Weyl) group $G_\Delta$ and for
any degenerate potentials. From this follows
\begin{equation}
i[{\cal H}_S,L_S^k]=[L_S^k,M_S],\quad \mbox{or} \quad
i[{\cal H}_S,(L_S^k)_{\mu\nu}]=\sum_{\kappa\in{\cal
R}}\left\{(L_S^k)_{\mu\kappa}(M_S)_{\kappa\nu}-
(M_S)_{\mu\kappa}(L_S^k)_{\kappa\nu}\right\}.
\label{spinLax2}
\end{equation}
Thanks to the sum up to zero condition of $M_S$ (\ref{sumMSzero}) we obtain
the conserved quantity as the {\em Total sum\/} (Ts) of $L_S^k$ instead of
the
diagonal sum (Tr):
\begin{equation}
[{\cal H}_S, \mbox{Ts}(L_S^k)]=0,\quad
\mbox{Ts}(L_S^k)\equiv\sum_{\mu,\nu\in{\cal R}}(L_S^k)_{\mu\nu}, \quad
k=2,\ldots,.
\label{totsum}
\end{equation}
This type of conserved quantities was known for the $A_r$ spin
Calogero-Moser models
for the vector representation \cite{suthsha2,HikWa}.
Note that (\ref{seciden2}) is  obtained from (\ref{seciden}) by replacing
$X$ and $\widetilde{M}$ by $X_S$ and $\widetilde{M}_S$.

Some remarks are in order.
\begin{enumerate}
\item
When all the spins are the same
\[
\psi_S^{(1)}=\psi_S^{(2)}=\cdots=\psi_S^{(D)}
\]
the action of the spin exchange operators become that of the identity
operator
\[
\hat{\cal P}_{\rho}=1,\quad \forall \rho\in\Delta.
\]
Then the Hamiltonian ${\cal H}_S$ (\ref{spinHamdef}) reduces to that of the
quantum Calogero-Moser models and the Lax operator
$L_S$ and $M_S$ become identical to the universal quantum
Lax pair operator derived
by Bordner, Manton and Sasaki \cite{bms}.
\item
The form of the spin Calogero-Moser Hamiltonian (\ref{spinHamdef}) depends
on the root system $\Delta$ only, although its actual operator contents
depend on the chosen representation ${\cal R}$.
\item
For the $A_r$ model with the vector representation, the present spin
Calogero-Moser coincides with the existing one. For the other root systems
the present model is completely new, to the best of our knowledge
(see the remarks
in the following entry).
It should be emphasised that even for the $A_r$ root system
the present formulation of the spin Calogero-Moser models defines
an infinitely many different models corresponding to the infinitely
many irreducible representations of the symmetric group ${\cal S}_{r+1}$,
which is the Weyl group of $A_r$.
\item
For the $A_r$ model with the vector representation the number of ``sites"
is $r+1$ which is equal to the degrees of freedom of the associated particle
motion, if the $A_r$ root system is embedded into ${\bf R}^{r+1}$ as is
done customarily.
This is a rather exceptional situation.
In all the other irreducible representations ${\cal S}_{r+1}$ and
for all the other root systems (except for the trivial
representation), the number of sites, or the dimensions
of ${\cal R}$, is bigger than $r$, the rank of $\Delta$.
For example, the vector representation of $D_r$ or the set of short roots
for $B_r$ consists of $2r$ vectors, which in a conventional parametrisation
of the roots take the form ${\cal R}=\{\pm {\bf e}_j, j=1,\ldots,r|
{\bf e}_j\in{\bf R}^r, {\bf e}_j\cdot{\bf e}_k=\delta_{jk}\}$.
Our spin Calogero-Moser models require all these $2r$ sites.
There are some references in which spin Calogero-Moser models for
$B_r$, $C_r$, $D_r$ or $BC_r$ are discussed \cite{simal,ber,yam}. In all
these papers, the number of sites is equal to the rank of the root systems.
These are different from the present spin Calogero-Moser models.
\item
The present formulation of the spin Calogero-Moser models together with the
Lax pair formulation does not require any specific structure of the ``spin"
space ${\bf V}$ attached to each site.
\item
It is well-known that for the spin 1/2 case in the $A_r$
model with the vector representation, the spin exchange operators
$\{\hat{\cal P}_\rho\}$ can be expressed in terms of the local
Pauli spin matrix at each site as
$\hat{\cal P}_{{\bf e}_j-{\bf
e}_k}=(1+\vec{\sigma}_j\cdot\vec{\sigma}_k)/2$.
For the vector representation of $D_r$ or
${\cal R}$ being the set of short roots
for $B_r$ mentioned above, we have
\begin{equation}
\hat{\cal P}_{{\bf e}_j}=[(1+\vec{\sigma}_j\cdot\vec{\sigma}_{-j})/2],
\quad
\begin{array}{l}
\hat{\cal P}_{{\bf e}_j-{\bf
e}_k}=[(1+\vec{\sigma}_j\cdot\vec{\sigma}_k)/2]
[(1+\vec{\sigma}_{-j}\cdot\vec{\sigma}_{-k})/2],\\[10pt]
\hat{\cal P}_{{\bf e}_j+{\bf
e}_k}=[(1+\vec{\sigma}_j\cdot\vec{\sigma}_{-k})/2]
[(1+\vec{\sigma}_{-j}\cdot\vec{\sigma}_{k})/2].
\end{array}
\end{equation}
In other words, $\hat{\cal P}_{{\bf e}_j+{\bf
e}_k}$ exchanges the spins at site $j$ and $-k$ and simultaneously the spins
at
$-j$ and $k$. Similar expressions
exist for other representations and root systems for $su(2)$, $su(N)$ or
other
spins.
\item
It is easy to verify, as in the Calogero-Moser models,
that the Hamiltonian ${\cal H}_S$ (\ref{spinHamdef}) is obtained as the
lowest member of the conserved quantities derived from the Lax
pair formulation:
\begin{equation}
{\cal H}_S\propto\mbox{Ts}(L_S^2).
\end{equation}
\item
The conserved quantities $\{\mbox{Ts}(L_S^k)\}$ are essentially the same as
those obtained in terms of the Dunkl \cite{Dunk} operators, and/or the
exchange
operator formalism \cite{fmp}. The same remark applies to the conserved
quantities of the spin exchange models to be discussed
in the following section.
For the quantum CM models without spin, the equivalence
of the Lax pair formalism
and Dunkl operator formalism was proven in
\cite{kps}.
\item The Yangian symmetry \cite{yangian,hikami} for the spin CM model and
spin
exchange model based on any root system is an interesting challenge.
\item
The commutativity of the conserved quantities obtained from the
above Lax pair formulation will be discussed elsewhere.

\end{enumerate}

\subsection{Rational Spin Calogero Model}
In this subsection we will define rational spin Calogero-Moser model
with quadratic confining potential to be called rational spin Calogero
model,
for brevity.
The Hamiltonian is given by
\begin{equation}
{\cal H}_{RS}=
{1\over 2} p^{2} + {1\over2}\omega^2q^2+{1\over2}\sum_{\rho\in\Delta_+}
   {|\rho|^{2}g_{|\rho|}(g_{|\rho|}-
\hat{\cal P}_\rho)\over{(\rho\cdot q)^2}}.
\label{ratCalHam}
\end{equation}
The construction of the Lax pair follows the same pattern as the case
without the spin degrees of freedom. Since the added potential
${1\over2}\omega^2q^2$ commutes with $X_S$, the canonical equations of
motion to be obtained from ${\cal H}_{RS}$ are equivalent to
\begin{equation}
\dot{L}_S=i[{\cal H}_{RS},L_S]=[L_S,M_S]-\omega^2Q,\quad
Q\equiv q\cdot\hat{H},
\label{spinLax3}
\end{equation}
in which $L_S$ and $M_S$ are the Lax pair for the rational ($1/(\rho\cdot
q)^2$) potential only.
Let us define
\begin{equation}
L_S^{\pm}=L_S\pm i\omega Q,
\end{equation}
whose time evolution read
\begin{equation}
\dot{L}_S^\pm
 = [L_S^\pm, M_S]\pm i\omega L^\pm.
\end{equation}
Here we have used well-known relations \cite{bcs2,bms}
\begin{equation}
\dot{Q}=p\cdot\hat{H}=L_S-X_S,\quad [Q,M_S]=-X_S.
\label{QMX}
\end{equation}
If we define
\begin{equation}
{\cal L}_S=L_S^+L_S^-,
\end{equation}
its time evolution is Lax like:
\begin{equation}
\dot{\cal L}_S=i[{\cal H}_{RS},{\cal L}_S]=[{\cal L}_S,M_S].
\end{equation}
Thus we obtain conserved quantities
\begin{equation}
\mbox{Ts}({\cal L}^k),\quad k=1,\ldots.
\end{equation}
The lowest conserved quantity Ts$({\cal L})$ gives the Hamiltonian
${\cal H}_{RS}$ (\ref{ratCalHam})
\begin{equation}
\mbox{Ts}({\cal L})\propto {\cal H}_{RS}+({r\over2}
+\sum_{\rho\in\Delta_+}
g_{|\rho|}\hat{\cal P}_\rho),
\end{equation}
plus additional terms which commute with all the spin exchange
operators $\{\hat{\cal P}_\rho\}$.

\section{Spin Exchange Model}
\label{spinexch}
\setcounter{equation}{0}

The spin exchange model is defined for a root system $\Delta$ and
an irreducible representation ${\cal R}$ of the reflection (Weyl) group
$G_\Delta$.
Its dynamical  state
is represented by a vector $\psi_S$ only which takes value in the $D$
multiple of a vector space ${\bf V}$;
\begin{equation}
\psi_S\in {\buildrel D\over\otimes}{\bf V}.
\end{equation}
As in the spin Calogero-Moser model case each $\bf V$ is associated with
site $\mu$. In other words
$\psi_S$ can be represented by its component spin $\psi_S^{(\mu)}$
at the site $\mu$, or $\psi_S^{(j)}$ at site $j$ for short:
\[
\psi_S=|\psi_S^{(1)},\ldots,\psi_S^{(D)}>.
\]
In fact, the spin exchange model is obtained from the corresponding
spin Calogero-Moser model by ``freezing" the particle degrees of freedom:
\begin{equation}
p=0,\quad q=q_0,
\end{equation}
in which $q_0$ is an equilibrium position of the classical
Calogero-Moser potential
\begin{equation}
\left.{\partial{\cal V}\over{\partial q}}\right|_{q=q_0}=0,
\qquad {\cal V}={1\over2}\sum_{\rho\in\Delta_+}
   {g_{|\rho|}^{2} |\rho|^{2}}
   \,V(\rho\cdot q).
\end{equation}
Since the rational potential without the quadratic confining potential or
the hyperbolic potential do not have any equilibrium points, this
automatically selects the trigonometric potential.
The rational potential with the quadratic confining potential case will be
discussed in the next subsection separately.
The equilibrium position $q_0$ for the trigonometric potential is
determined uniquely in each Weyl alcove.
In other words, if $q_0$ is an equilibrium point so is
$s_{\alpha}(q_0)$ which
defines an equally integrable model.
 Let us fix $q_0$ and define $X_E$ and
$\widetilde{M}_E$ in terms of the Lax pair operators of
the corresponding spin
Calogero-Moser model at $q=q_0$:
\begin{equation}
X_E=X_S|_{q=q_0},\quad \widetilde{M}_E=\widetilde{M}_S|_{q=q_0}.
\end{equation}
The components of the matrices $X_E$ and $\widetilde{M}_E$ are
linear combinations of the spin exchange operators $\hat{\cal P}_\rho$
and the coefficients are just numbers. They satisfy a simple matrix identity
\begin{equation}
[X_E,\widetilde{M}_E]=0
\label{Eiden}
\end{equation}
and as before $\widetilde{M}_E$ has a special property:
\[
   \sum_{\mu\in{\cal R}}(\widetilde{M}_E)_{\mu\nu}=
   \sum_{\nu\in{\cal R}}(\widetilde{M}_E)_{\mu\nu}=
   -i{\cal A}_E,\quad
{\cal A}_E={1\over2}\sum_{\rho\in\Delta_{+}}g_{|\rho|}|\rho|^2\,V
      (\rho\cdot q_0)\hat{\cal P}_\rho,
\]
As in the previous section we  define a new matrix $M_E$,
\[
   M_E=\widetilde{M}_E+i{\cal A}_E\times I,
\]
which satisfies {\em sum up to zero} condition, too
\begin{equation}
   \sum_{\mu\in{\cal R}}({M_E})_{\mu\nu}=
   \sum_{\nu\in{\cal R}}({M_E})_{\mu\nu}=0.
   \label{sumMEzero}
\end{equation}
By rewriting (\ref{Eiden}) in terms of $M_E$ we arrive at the
Lax representation of the spin exchange model:
\begin{equation}
i[{\cal H}_E,X_E]=[X_E,M_E],
\end{equation}
in which the Hamiltonian ${\cal H}_E$ of the spin exchange model is
\begin{equation}
{\cal H}_E={1\over2}\sum_{\rho\in\Delta_{+}}g_{|\rho|}|\rho|^2\,V
      (\rho\cdot q_0)(1-\hat{\cal P}_\rho)=-{\cal A}_E + const.
\label{spinEHam}
\end{equation}
The added constant simply shifts the ground state energy.
The Lax pair supplies the conserved quantities  as the
{\em Total sum} of $X_E^k$:
\begin{equation}
[{\cal H}_E, \mbox{Ts}(X_E^k)]=0,\quad
\mbox{Ts}(X_E^k)\equiv\sum_{\mu,\nu\in{\cal
R}}(X_E^k)_{\mu\nu},
\quad k=3,\ldots,.
\label{totsum2}
\end{equation}
It is interesting to note that the first two members Ts$(X_E^1)$ and
Ts$(X_E^2)$ are trivial, in contrast to the spin Calogero-Moser case.

Some remarks are in order.
\begin{enumerate}
\item
As in the spin Calogero-Moser model,
the form of the spin exchange model Hamiltonian ${\cal
H}_E$ (\ref{spinEHam}) depends on the root system $\Delta$ only, although
its actual operator contents depend on the chosen representation ${\cal
R}$. The infinitely many models corresponding to various irreducible
representations, sharing the same set of conserved quantities, can be
considered to constitute an integrable {\em hierarchy\/} belonging to the
root system $\Delta$.
If one considers a series of representations with increasing
dimensionality ({\em i.e.\/} more spins),
the thermodynamic limit could be achieved within models
belonging to a fixed root system $\Delta$. This is a novel
situation, since in the Haldane-Shastry model the rank $r$
grows  indefinitely in the thermodynamic
limit.
\item
It should be remarked that the $q_0$ is the equilibrium point {\em not}
of the function appearing in the Hamiltonian ${\cal
H}_E$ (\ref{spinEHam}) which is linear in the coupling constants
$g_{|\rho|}$ but that of the potential of the {\em classical} Calogero-Moser
Hamiltonian ${\cal H}_C$ (\ref{cCMHamiltonian}) which is quadratic in the
coupling constants. This difference is meaningful only for the models
based on non-simply laced root systems.
\item
It should be emphasised that the ``coordinates" $q$
or rather $q_0$ are just a set
of numbers rather than dynamical variables.
Thus, in contrast to the conventional approach \cite{halsha,fmp}, the notion
of `position exchange operator' is not used in our approach.
\item
For the $A_r$ model, $q_0$ can be chosen to be ``equidistant":
\begin{equation}
q_0=\pi(1,2,\ldots,r,r+1)/(r+1),
\label{equi}
\end{equation}
thanks to the well-known trigonometric identity
\[
\sum_{k\neq j}^{r+1}{\cos\,[\pi(j-k)/(r+1)]\over{\sin^3[\pi(j-k)/(r+1)]}}=0.
\]
The Haldane-Shastry model \cite{halsha}, {\em i.e.} the $A_r$ spin exchange
model for the vector representation, has been understood quite well because
of this simplifying feature.
\item
The equidistance of  $q_0$ for $A_r$ seems rather fortuitous.
As remarked above,
any transposition of the above $q_0$ (\ref{equi}) provides an
equally integrable spin exchange model,
but the equidistance property is lost.
As for $D_r$ ($r\ge4$), we have not been able to find equidistant $q_0$.
For
$BC_r$ model, equidistant
$q_0$ can be achieved  for certain ratios of the coupling constants.
For the following parametrisation of the potential \cite{Ino0,bcs1},
\begin{eqnarray}
{\cal V}=\sum_{j<k}^r
   \left[{{g_{M}^{2} }\over{\sin^2(q_j-q_k)}}+{{g_{M}^{2}
}\over{\sin^2(q_j+q_k)}}\right]+ \sum_{j=1}^r{{g_{S}(g_{S}+2g_{L})
}\over{2\sin^2(q_j)}}+\sum_{j=1}^r{2{g_{L}^{2} }\over{\sin^2(2q_j)}},
\end{eqnarray}
one obtains equidistant equilibrium positions:
\begin{eqnarray}
q_0&=&\pi(1,3,\ldots,2r-1)/4r,\quad
\mbox{for}\quad g_L/g_M=1/2,\quad g_S=0,\\
q_0&=&\pi(1,2,\ldots,r)/2(r+1),\quad
\mbox{for}\quad g_L/g_M=3/2,\quad g_S=0,\\
q_0&=&\pi(1,2,\ldots,r)/(2r+1),\quad
\mbox{for}\quad g_L/g_M=1/2,\quad g_S/g_M=1.
\end{eqnarray}
These cases were discussed in some detail
by Bernard-Pasquier-Serban \cite{ber}.
\item
Note that the present derivation of the spin exchange model and its
Lax pair does not adopt the strong coupling limit.
\item
For most general elliptic potentials, the Lax pair can be constructed
in a usual manner \cite{OP1}. But the second Lax operator does not satisfy
the
``sum to zero" condition, hence the integrability of these models is not yet
established.
\end{enumerate}

\subsection{Rational Spin Exchange Model}
The above formulation fails to give integrable spin exchange model with
rational potential. This can be remedied by adding a harmonic confining
potential \cite{fmp,yam}
which creates equilibrium points in each Weyl chamber. Here we derive the
Lax operator formalism for these models. Let us start with the Lax pair
for the rational Calogero-Moser models and for the time being keep the
value of $q$ unspecified.
We have as in (\ref{firstiden})
\[
[X,\widetilde{M}]=-\hat{H}\cdot{\partial{\cal V}\over{\partial q}}
\]
and after multiplying $\hat{\cal P}_\rho$ to functions $x(\rho\cdot q)$ and
$y(\rho\cdot q)$, we obtain (\ref{mainiden})
 \begin{equation}
[X_S, \widetilde{M}_S]=-\hat{H}\cdot{\partial{\cal V}\over{\partial q}}.
\label{xmsnew}
\end{equation}
The diagonal matrix $Q$ (\ref{spinLax2}) satisfies the relation (\ref{QMX})
\begin{equation}
[Q,\widetilde{M}_S]=-X_S.
\label{qmsnew}
\end{equation}
If we define two new matrices $X_S^\pm$
\begin{equation}
X_S^\pm=X_S\pm i\omega Q,
\end{equation}
they satisfy simple commutation relations thanks to (\ref{xmsnew}) and
(\ref{qmsnew})
\begin{equation}
[X_S^\pm,\widetilde{M}_S]=\mp i\omega X_S^\pm-\hat{H}\cdot{\partial{\cal
V}_{RC}\over{\partial q}},
\end{equation}
in which ${\cal V}_{RC}$ is the potential
of the classical rational Calogero-Moser model with harmonic confining
potential
\begin{equation}
{\cal V}_{RC}={1\over2}\omega^2q^2+{1\over2}\sum_{\rho\in\Delta_+}
   {g_{|\rho|}^2|\rho|^{2}\over{(\rho\cdot q)^2}}.
\end{equation}
Now we choose $q_0$ as an equilibrium point of ${\cal V}_{RC}$ and define
\begin{equation}
X_{RE}^\pm=X_S^\pm|_{q=q_0},\quad
\widetilde{M}_{RE}=\widetilde{M}_S|_{q=q_0},
\qquad
\left.{\partial{\cal V}_{RC}\over{\partial q}}\right|_{q=q_0}=0.
\end{equation}
Thus we arrive at
\begin{equation}
[X_{RE}^+X_{RE}^-,\widetilde{M}_{RE}]=X_{RE}^+[X_{RE}^-,\widetilde{M}_{RE}]
+[X_{RE}^+,\widetilde{M}_{RE}]X_{RE}^-=0.
\label{XXMzero}
\end{equation}
We define $M_{RE}$ by
\begin{equation}
   M_{RE}=\widetilde{M}_{RE}+i{\cal A}_{RE}\times I,
   \label{MREdefnew}
\end{equation}
\begin{equation}
{\cal A}_{RE}={1\over2}\sum_{\rho\in\Delta_{+}}{g_{|\rho|}|\rho|^2
\hat{\cal P}_\rho\over{(\rho\cdot q_0)^2}},
\end{equation}
so that $M_{RE}$ satisfies the {\em sum  to zero} condition
\begin{equation}
   \sum_{\mu\in{\cal R}}({M_{RE}})_{\mu\nu}=
   \sum_{\nu\in{\cal R}}({M_{RE}})_{\mu\nu}=0.
   \label{sumMREzero}
\end{equation}
Then (\ref{XXMzero}) can be rewritten as a Lax representation for the
rational spin exchange model
\begin{equation}
i[{\cal H}_{RE}, X_{RE}^+X_{RE}^-]=[X_{RE}^+X_{RE}^-,M_{RE}],
\end{equation}
in which the
rational spin exchange Hamiltonian ${\cal H}_{RE}$ is defined by
\begin{equation}
{\cal H}_{RE}={1\over2}\sum_{\rho\in\Delta_{+}}{g_{|\rho|}|\rho|^2\over{
      (\rho\cdot q_0)^2}}(1-\hat{\cal P}_\rho)
-{\cal A}_{RE} + const.
\label{spinREHam}
\end{equation}
The conserved quantities are obtained as
{\em Total sum} of $(X_{RE}^+X_{RE}^-)^k$:
\begin{equation}
\left[{\cal H}_{RE},
\mbox{Ts}\left(({X_{RE}^+X_{RE}^-)}^k\right)\right]=0,\quad
 \quad k=1,\ldots,.
\label{totsumRE}
\end{equation}
It is interesting to note that the above Hamiltonian ${\cal H}_{RE}$
depends on the harmonic confining potential ${1\over2}\omega^2q^2$
only through the value $q_0$.

\section{Summary and comments}
\label{comdis}
\setcounter{equation}{0}

We have shown that the integrability of spin Calogero-Moser model and
the spin exchange model with degenerate potential and based
on any root system is a
direct consequence of the integrability
of the corresponding classical Calogero-Moser system.
For a given root system $\Delta$ there are infinitely many integrable
spin Calogero-Moser models and
the spin exchange models corresponding to infinitely many irreducible
representations ${\cal R}$ of the reflection group.
These define  physically different models sharing the same exchange
features.

After completion of the present work, we came across
\cite{spaingr} which discusses
the integrability of
spin $BC_r$ model with harmonic confining potential,
or ``spin Inozemtsev model"
\cite{Ino0} in terms of the Dunkl operator formalism \cite{Dunk}.

\section*{Acknowledgements}
\setcounter{equation}{0}
We thank D.\,B.\,Fairlie for bringing \cite{spaingr} to our attention.
V.\,I.\,I. is supported by JSPS long term fellowship.
R. S. is partially
supported  by the Grant-in-aid from the Ministry of Education, Culture,
Sports, Science and Technology, Japan,  priority area (\#707)
``Supersymmetry and unified theory of elementary particles".

\end{document}